\documentclass[11pt,twoside,a4paper,cmspaper,final,collab]{cms-tdr}
\def\svnVersion{4c3cc7c}\def\svnDate{2026/03/13}\def\cmsCernNoTag{CERN-EP-2025-112}\def\cmsCernDate{\today}\def\cmsMessage{Submitted to Physics Letters B}
\begin{document}\cmsNoteHeader{HIN-24-008}

\newcommand{\PbPb}{\ensuremath{\text{PbPb}}\xspace}
\newcommand{\pp}{\ensuremath{\Pp\Pp}\xspace}
\newcommand{\pTtrig}{\ensuremath{p_\text{T,trig}}\xspace}
\newcommand{\pTasso}{\ensuremath{p_\text{T,asso}}\xspace}
\newcommand{\Etatrig}{\ensuremath{\eta_\text{trig}}\xspace}
\newcommand{\Etaasso}{\ensuremath{\eta_\text{asso}}\xspace}
\newcommand{\PYQUEN}{\textsc{pyquen}\xspace}
\newlength\cmsTabSkip\setlength{\cmsTabSkip}{1ex}

\ifthenelse{\boolean{cms@external}}{\providecommand{\cmsTable}[1]{\resizebox{\columnwidth}{!}{#1}}}{\providecommand{\cmsTable}[1]{#1}}
\ifthenelse{\boolean{cms@external}}{\providecommand{\cmsLeftUpperCase}{Upper\xspace}}{\providecommand{\cmsLeftUpperCase}{Left\xspace}}
\ifthenelse{\boolean{cms@external}}{\providecommand{\cmsLeft}{upper\xspace}}{\providecommand{\cmsLeft}{left\xspace}}
\ifthenelse{\boolean{cms@external}}{\providecommand{\cmsRight}{lower\xspace}}{\providecommand{\cmsRight}{right\xspace}}

\cmsNoteHeader{HIN-24-008}

\title{
Jet peak shapes based on two-particle angular correlations in lead-lead collisions at \texorpdfstring{$\sqrt{{s_{_{\mathrm{NN}}}}} = 5.02\TeV$}{sNN = 5.02 TeV}
}

\date{\today}

\abstract{
The longitudinal invariance of jet-induced peaks in two-particle correlation functions from relativistic lead-lead collisions is experimentally explored. The data were collected at a center-of-mass energy per nucleon pair of 5.02\TeV in 2018 using the CMS detector. The dataset corresponds to an integrated luminosity of 0.607\nbinv. Long- and short-range correlations are studied through two-dimensional distributions of the separations in pseudorapidity and azimuth between particles in an event. Jets manifest as a well-defined peak at small angular separations, and the shape of this peak provides insight into jet medium interactions. This Letter examines the evolution of the jet peak shape, focusing on the dependence of its width and longitudinal asymmetry on the transverse momentum, collision centrality, and pseudorapidity of the associated charged particles. The jet-peak distributions of lower transverse momentum particles broaden in both pseudorapidity and azimuth with increasing collision overlap, with the broadening in pseudorapidity being more pronounced. The longitudinal asymmetry of the peaks is also found to increase as the average pseudorapidity increases. These results are compared to proton-proton collision data that were obtained at the same nucleon-nucleon collision center-of-mass energy with an integrated luminosity of 252\nbinv.
}

\hypersetup{
pdfauthor={CMS Collaboration},
pdftitle={Jet peak shapes based on two-particle angular correlations in lead-lead collisions at sqrt(sNN) = 5.02 TeV},
pdfsubject={CMS},
pdfkeywords={CMS, heavy ion, two-particle correlation, longitudinal asymmetry}
}
\maketitle 

\section{Introduction}
\label{sec:intro}
 
In ultrarelativistic heavy ion collisions, a deconfined state of quarks and gluons, known as the quark-gluon plasma (QGP), 
forms as the system reaches equilibrium~\cite{Theory_qgp, Theory_HIN}.
In the early stage of such collisions, high-momentum transfer (hard) processes result in high-energy partons. These lead to collimated streams of particles (jets) that interact with the QGP through ``jet quenching'' phenomena~\cite{Theory_jetquecnche}.
Experimental evidence for jet quenching has been observed at both the BNL RHIC~\cite{RHIC_jetquecnche_1, RHIC_jetquecnche_2} and the CERN LHC~\cite{LHC_jetquecnche_1, LHC_jetquecnche_2, LHC_jetquecnche_3, LHC_jetquecnche_4}.

While many previous studies of jet quenching have focused on reconstructing the final-state particle distributions associated with jets at high transverse momentum (\pt)~\cite{CMS_FR_2}, two-particle correlations in pseudorapidity ($\eta$) and azimuth ($\phi$) offer a complementary method for studying jet properties in heavy ion collisions. These correlations measure the relative separations in pseudorapidity ($\Delta\eta$) and azimuth ($\Delta\phi$)  of a high-\pt ``trigger'' particle and lower-\pt associated particles. Jets appear as a well-defined ``near-side'' peak at ($\Delta\eta = 0$, $\Delta\phi = 0$). In these correlations, an elongated ``away-side'' structure in $\Delta\eta$ at $\Delta\phi = \pi$ is also typically present, corresponding to the recoil parton in a dijet event. For specific kinematic selections of trigger and associated particle transverse momenta, heavy ion collisions reveal a suppression of the away-side and a moderate enhancement for the near-side correlation yields in collisions where there is a large overlap of the nuclei, indicating medium-induced modifications~\cite{CMS_Jet_yld_EURO, CMS_FR_1, CMS_FR_3, CMS_Jet_charge_JHEP}. The shape of the jet peak arising from particle correlations within a given jet cone is also sensitive to medium-induced modifications. Two-particle correlation  measurements of jet peak shapes allow for simple background corrections and cover a broader range of jet energies and early stage dynamics than typically explored in studies with fully reconstructed jets~\cite{ALI_Jet_yld_PRL}.

In high-energy heavy ion collisions, an approximate invariance is expected of jet yields under longitudinal boosts near midrapidity~\cite{Srivastava:1992cg}. However, at forward rapidities, this invariance may break down as a result of the longitudinal expansion of the created medium and the evolving parton densities. Particle densities within given jets can also be affected by the longitudinal expansion. In this study, the shapes of near-side jet peaks in two-particle angular correlations are used to characterize the particle densities within high-\pt jets. By measuring this observable over a wide $\eta$ range, we investigate the extent to which longitudinal invariance holds and how longitudinal dynamics modify the jet-associated yields. These measurements also provide sensitivity to kinematic regions where initial-state effects, such as parton saturation~\cite{parton_saturation}, may become increasingly relevant.

Using PbPb collision data, this Letter examines the angular distribution of associated charged particles relative to a charged hadron trigger across the full $\phi$ and wide $\eta$ coverage ($\abs{\eta} < 2.4$) of the CMS detector at the LHC. The data are collected at a center-of-mass energy per nucleon pair of $\sqrtsNN = 5.02\TeV$. Results are also reported for proton-proton (\pp) collisions at the same nucleon-nucleon collision center-of-mass energy to provide a reference. Using two-particle correlations, the longitudinal ($\Delta\eta$) and transverse ($\Delta\phi$) widths of the near-side jet peak are determined. In addition, the longitudinal invariance of this peak shape is studied for higher-momentum trigger particles at forward rapidities. 

Tabulated results are provided in the HEPData record for this analysis~\cite{hepdata}.

\section{Experimental setup and data sample}
\label{sec:exp}

The central feature of the CMS apparatus is a superconducting solenoid of 6\unit{m} internal diameter, providing a magnetic field of 3.8\unit{T}. Within the solenoid volume, there is a silicon pixel and strip tracker that measures the trajectories of charged particles in the range $\abs{\eta}<2.5$, a lead-tungstate crystal electromagnetic calorimeter, and a brass and scintillator hadron calorimeter. For charged particles of $1<\pt<10\GeV$ and $\abs{\eta}<1.4$, the typical track resolution is 1.5\% in \pt and 25--90 (45--150)\micron in the transverse (longitudinal) impact parameter~\cite{CMS:2014pgm, DP-2020-049, Phase1Pixel}. Steel and quartz-fiber Cherenkov hadron forward (HF) calorimeters cover the range $2.9 < \abs{\eta} < 5.2$ on either side of the interaction region and are azimuthally divided into $20^\circ$ modular wedges, which are further segmented to form $0.175 \times 0.175 \unit{radians}$ ($\Delta\eta \times \Delta\phi$) towers. For heavy ion collisions, centrality is defined as the fraction of the total inelastic hadronic cross section, with 0\% corresponding to the full overlap of the two colliding nuclei. A detailed description of the CMS experiment can be found in Ref.~\cite{CMS:2008xjf, CMS:2024_run3}.

The analysis is based on \PbPb (\pp) collisions at $\sqrtsNN= 5.02\TeV$ data, which were collected by the CMS experiment in 2018 (2017) with a two-tiered trigger system. The first level trigger consists of custom hardware processors and uses information from the calorimeters and muon detectors to select events at a rate of around 100\unit{kHz}~\cite{CMS:2020cmk}. The second level or high-level trigger consists of a farm of processors running a version of the full event reconstruction software optimized for fast processing. The \PbPb and \pp data sets correspond to integrated luminosities of 0.607\nbinv~\cite{CMS:LUM-18-001} and 252\nbinv~\cite{CMS:LUM-17-003,CMS:LUM-17-004,CMS:LUM-18-002}, respectively, with approximately 4.3 billion \PbPb minimum-bias (MB) events and 15 billion \pp MB events.The MB events are selected by requiring energy deposition above thresholds in the range of approximately 6--12\GeV in both of the HF calorimeters located on either side of the interaction region~\cite{CMS:2016ngn}. Further selections are applied offline to reject events from background processes (beam-gas interactions and non-hadronic collisions), as discussed in Ref.~\cite{strangehadron}. In the offline analysis, events are required to have at least one interaction vertex, based on two or more reconstructed tracks, with a distance of less than 15\unit{cm} from the center of the nominal interaction point along the beam axis, $v_\text{z}$. The primary vertex (PV) is defined as the vertex with the highest sum of the squares of the \pt values of its associated tracks in the event, as described in Section 9.4.1 of Ref.~\cite{CMS-TDR-15-02}. In the final analysis, the \PbPb collision events are required to have at least two calorimeter towers in the HF detectors on both sides of the interaction region with energy deposits of more than 4\GeV per tower.  These criteria remove approximately $99$\% of ultraperipheral (non-hadronic) collision events.  

The selected tracks are corrected for tracking efficiency and acceptance using simulated Monte Carlo samples produced by \PYTHIA~8.212~\cite{Sjostrand:2014zea} with the CP5 tune~\cite{Sirunyan:2019dfx} for \pp collisions, and \HYDJET 1.9~\cite{Lokhtin:2005px} for \PbPb collisions. The \HYDJET generator simulates both soft and hard components in heavy ion collisions by combining a parametrized hydrodynamic expansion with multiple \PYTHIA/{\PYQUEN} events. While the soft component (track $\pt \lesssim 5\GeV$) in these simulations shows approximate boost invariance, including at forward rapidity, the hard component (track $\pt \gtrsim 5\GeV$) does not. This difference between the approximately boost-invariant soft component and the non-invariant hard component reflects \HYDJET's internal modeling assumptions rather than robust physical predictions. The response of the CMS detector to these simulated events is modeled using \GEANTfour~\cite{GEANT4}.

\section{Analysis procedure}
\label{sec:ana}

High-purity~\cite{CMS:2014pgm} tracks resulting from the passage of charged particles through the detector are selected in the range $\abs{\eta} < 2.4$, requiring that the significances of the reconstructed longitudinal distance ($d_\text{z}$) and the transverse distance ($d_\text{xy}$) from the beam axis satisfy $\abs{d_\text{z}}/\sigma_\text{z} < 3$ and $\abs{d_\text{xy}}/\sigma_\text{xy} < 3$, respectively, where $\sigma_\text{z}$ and $\sigma_\text{xy}$ denote the associated uncertainties in the measurements of $d_\text{z}$ and $d_\text{xy}$. Additionally, the \pt resolution for each track, expressed as $\sigma_\text{\pt}/\pt$, must be less than 10\%. The number of layers hit in the silicon tracker must be greater than ten, and the $\chi^{2}$ of the track fit per degree of freedom per layers hit in the silicon tracker is required to be less than 0.18. These conditions define the nominal track selection criteria.

\subsection{Longitudinal and transverse widths of the near-side jet peak}
\label{sec:ana1}

The correlation between two charged particles is calculated as a function of $\Delta\eta = \Etaasso-\Etatrig$ and $\Delta\phi = \phi_\text{asso} - \phi_\text{trig}$ (defined between $-\pi/2$ and $3\pi/2$ to visualize the back-to-back jet peaks). The correlation is expressed in terms of the associated yield per trigger particle in a given interval of trigger and associated particle \pt. The number of trigger particles for each interval of an event is denoted by $N_\text{trig}$, which is corrected for the tracking efficiency and the rate of misreconstructed tracks. 

The signal distribution, $S(\Delta \eta,\Delta \phi)$, is constructed using pairs of particles from the same event, and is normalized by the number of trigger particles. It is defined as 
\begin{equation}
S(\Delta \eta,\Delta \phi) = \frac{1}{N_\text{trig}}\frac{\rd^{2}N^\text{same}}{\rd{\Delta \eta} \,\rd{\Delta \phi}}
\label{equ:sig}
\end{equation}
where $N^\text{same}$ is the number of particle pairs in the ($\Delta \eta$, $\Delta \phi$) bin for the event~\cite{CMS_gamma-p_2pc, CMS_PLB_2013, CMS_PRL_2016}.

The mixed events distribution $B(\Delta \eta, \Delta \phi)$ is constructed by pairing the trigger particles in each interval of an event with the particles from ten different events~\cite{CMS_gamma-p_2pc, CMS_PLB_2013, CMS_PRL_2016}. The mixed-events are selected such that they are within the same centrality (in \PbPb collisions) or track multiplicity (in \pp collisions) range as the primary event, and such that the relative difference of the PV along the z-axis is less than 2\cm. The mixed-event distribution, $B(\Delta \eta, \Delta \phi)$, is given by
\begin{equation}
B(\Delta \eta,\Delta \phi) = \frac{1}{N_\text{trig}}\frac{\rd^{2}N^\text{mix}}{\rd{\Delta \eta} \,\rd{\Delta \phi}}
\label{equ:bkg}
\end{equation}
where $N^\text{mix}$ is the number of mixed-event pairs in a given ($\Delta\eta, \Delta \phi$) bin~\cite{CMS_Jet_yld_EURO, CMS_gamma-p_2pc, ALI_Jet_PRC, STAR_Jet_PRC, CMS_PLB_2013, CMS_PRL_2016}. The signal and mixed-event distributions are first calculated for each event, and then averaged over all events within the same centrality or track multiplicity class. The correlation function, also referred to as the per-trigger associated yield, is defined as
\begin{equation}
\frac{1}{N_\text{trig}}\frac{\rd^{2}N^\text{pair}}{\rd{\Delta \eta} \,\rd{\Delta \phi}} = C_{2}(\Delta \eta,\Delta \phi)=B(0,0)\frac{S(\Delta \eta,\Delta \phi)}{B(\Delta \eta,\Delta \phi)}
\label{equ:corrfun}
\end{equation}
where $N^\text{pair}$ is the pair-acceptance corrected same-event pairs in a given ($\Delta\eta, \Delta\phi$) bin.

\begin{figure*}[htbp]
    \centering
    \includegraphics[width=0.95\textwidth]{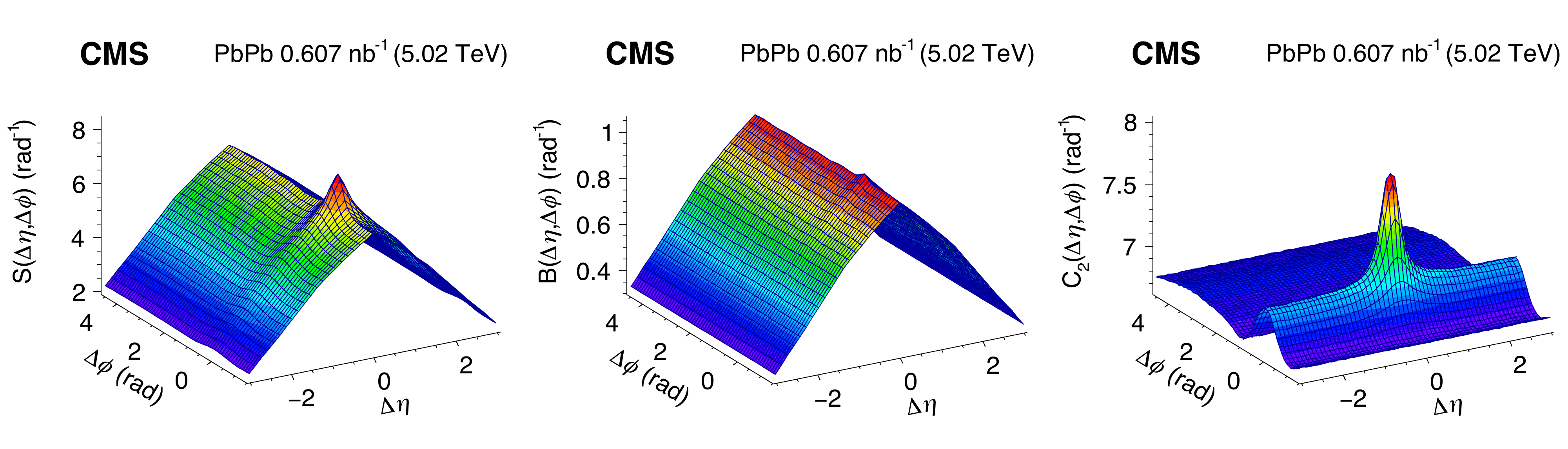}
    \caption{Average signal (left), mixed-event (middle), and pair-acceptance corrected per-trigger associated yield (right) distributions for 0--10\% centrality events, with $4 < \pTtrig < 8\GeV$ and $2 < \pTasso < 3\GeV$, in \PbPb collision data at 5.02\TeV.}
    \label{fig:sigbkgcorrfun_010}
\end{figure*}

Figure~\ref{fig:sigbkgcorrfun_010} (left) shows a typical signal distribution, which has the physical correlations in the form of a peak on the top of a triangular-shaped continuum. The mixed-event distribution, shown in Fig.~\ref{fig:sigbkgcorrfun_010} (middle), reflects this continuum, which results from detector-related artifacts, such as the pair-acceptance effect that is also present in the signal distribution. The ratio $\frac {B(0,0)} {B(\Delta \eta,\Delta \phi)}$ corrects for the pair-acceptance effects present in the signal distribution. After applying this correction, a residual $\Delta\eta$-dependent modulation can remain in the mixed-event corrected correlation function. Possible imperfections in the event mixing are examined by studying the $\Delta\eta$ distribution in the $\Delta\phi$ sideband region $1.2 < \Delta\phi < \pi - 1.2$. Any residual non-uniformity observed in this region is corrected by flattening the corresponding $\Delta\eta$ dependence. The final pair-acceptance corrected per-trigger associated yield distribution, after all corrections, is shown in Fig.~\ref{fig:sigbkgcorrfun_010} (right)~\cite{CMS_Jet_yld_EURO, CMS_gamma-p_2pc, ALI_Jet_PRC, STAR_Jet_PRC, CMS_PLB_2013, CMS_PRL_2016}.

A long-range background contribution remains in the distribution. This background is determined using the distribution with $\abs{\Delta\eta} > 2$ and is subtracted to isolate the near-side, short-range jet correlations.

\begin{figure}[htbp]
    \centering    
    \includegraphics[width=0.48\textwidth]{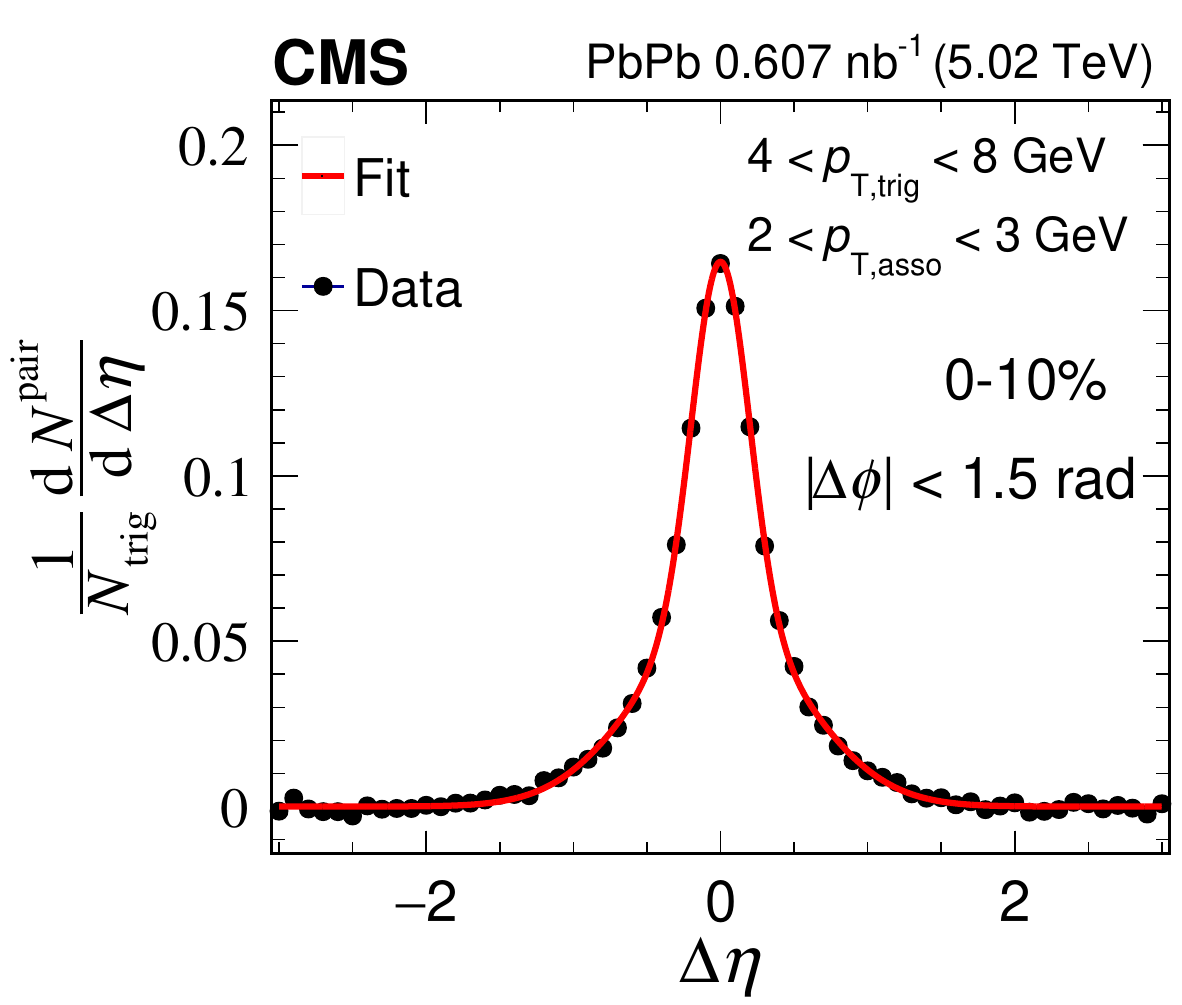}
    \includegraphics[width=0.48\textwidth]{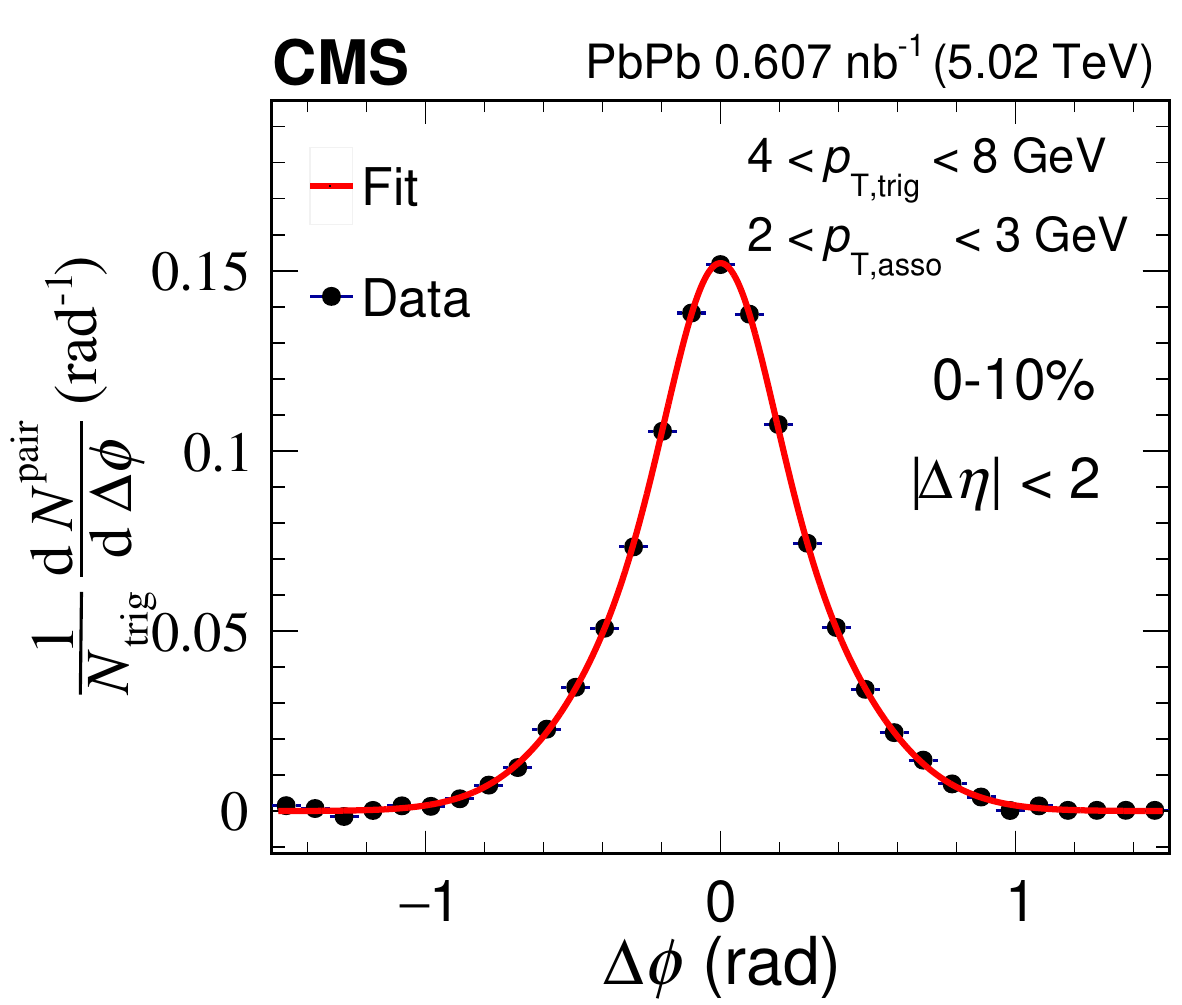}
    \caption{(\cmsLeftUpperCase) One-dimensional $\Delta\eta$ projection and (\cmsRight) One-dimensional $\Delta\phi$ projection.}
    \label{fig:1D_projection}
\end{figure}

To determine the longitudinal and transverse widths of the near-side jet peak, the projections of the two-dimensional (2D) correlations onto the $\Delta\eta$ and $\Delta\phi$ axes are fitted separately using a double-Gaussian function. The effective width ($\sigma$) of each projected distribution is defined as the amplitude-weighted quadratic mean of the two Gaussian widths of the respective distribution. Figure~\ref{fig:1D_projection} shows the one-dimensional (1D) $\Delta\eta$ and $\Delta\phi$ projections along with the associated fits for the respective ranges of $\abs{\Delta\eta} < 2$ and $\abs{\Delta\phi} < 1.5$.

\subsection{Longitudinal asymmetry of the near-side peak from mid to forward rapidity}
\label{sec:ana2}

The Longitudinal invariance of the near-side peak is investigated by studying the asymmetry of the associated yield around $\Delta\eta = 0$ as a function of $\Etatrig$ using two-particle correlations. The analysis is performed for eight trigger particle $\eta$ ranges, with range boundaries [$-2, -1.5, -1, -0.5, 0, 0.5, 1, 1.5, 2$]. For symmetric systems there should be mirror symmetry, with the distribution for $\Etatrig > 0$ being the mirror image of the distribution for $\Etatrig < 0$.

\begin{figure}[htbp]
    \centering
    \includegraphics[width=0.48\textwidth]{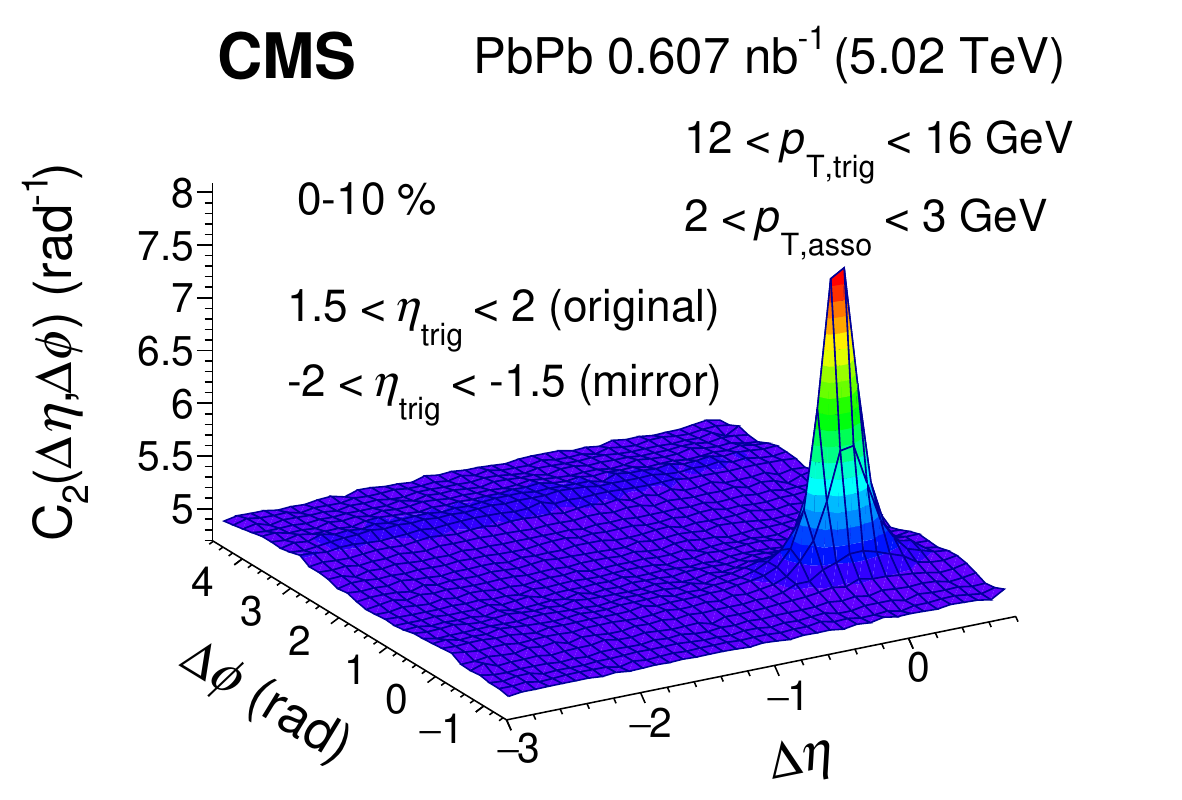}
    \includegraphics[width=0.48\textwidth]{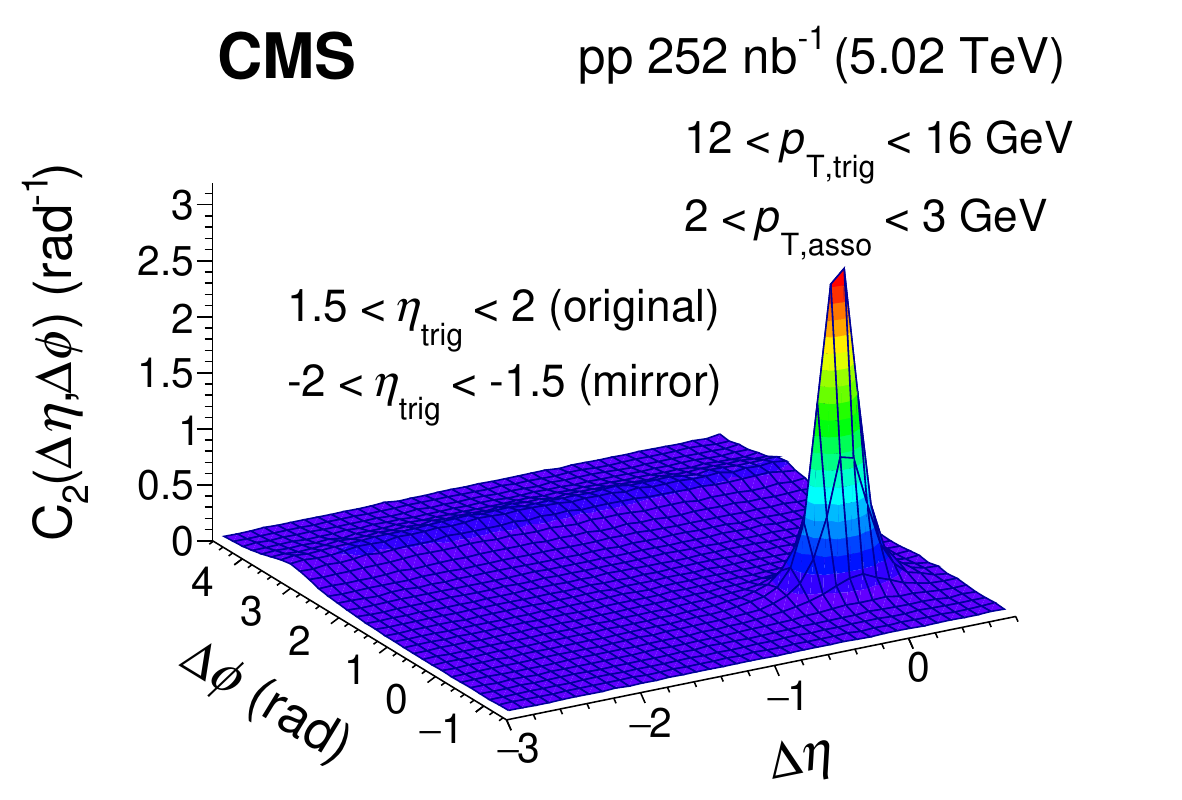}
    \caption{Two-dimensional correlation distributions in the 0--10\% centrality interval for \PbPb (\cmsLeft) and \pp (\cmsRight) collisions are shown, with $1.5 < \abs{\Etatrig} < 2$, $12 < \pTtrig < 16\GeV$, and $2 < \pTasso < 3\GeV$.}
    \label{fig:2Dcorr}
\end{figure}

Figure~\ref{fig:2Dcorr} shows the average of the $C_{2}(\Delta\eta, \Delta\phi)$ distributions for $1.5 < \Etatrig < 2$ and the mirror image of $C_{2}$ for $-2 < \Etatrig < -1.5$, with $\abs{\Etaasso} < 2.4$, in \PbPb (\cmsLeft) and \pp (\cmsRight) collisions. The \pt ranges for the trigger and associated particles are $12 < \pTtrig < 16\GeV$ and $2 < \pTasso < 3\GeV$, respectively. 

\begin{figure}[htbp]
    \centering
    \includegraphics[width=0.48\textwidth]{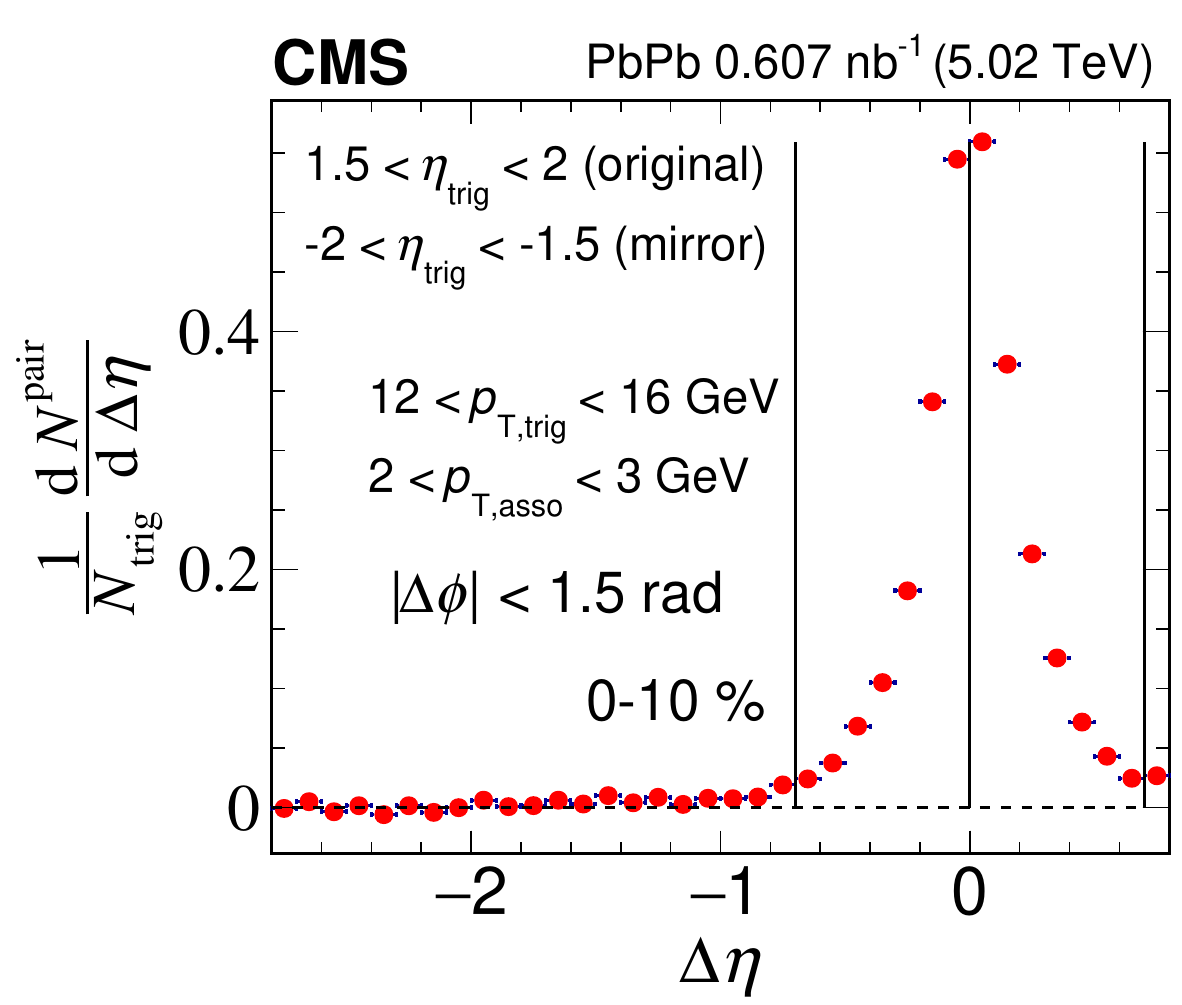}
    \caption{One-dimensional $\Delta\eta$ projection from the long-range background-subtracted averaged 2D correlation, combining $1.5 < \Etatrig < 2$ (original) and $-2 < \Etatrig < -1.5$ (mirror). Three vertical lines correspond to a $\Delta\eta$ value of --0.7, 0 and 0.7 from left to right. The dashed line indicates the zero level on the y-axis.}
    \label{fig:1Dyield}
\end{figure}

Fig.~\ref{fig:1Dyield} shows the 1D projections in $\Delta\eta$ with $\abs{\Delta\phi} < 1.5$ after subtracting a long-range background contribution. This long-range contribution is obtained using the projected values with $\abs{\Delta\eta} > 2$. The asymmetry of the jet peak shape is characterized by taking the ratio of associated yields ($Y$) within an $\Etaasso$ range of 0.7 units on either side of the peak, $Y_{\Delta\eta>0}/Y_{\Delta\eta<0}$. The range boundaries are chosen to avoid edge effects influencing the results in any of the measured $\Etatrig$ ranges.

\section{Systematic uncertainties}
\label{sec:syst}  

Systematic uncertainties arising from the selected vertex range, the track selection, the centrality calibration, tracking efficiency, pileup, pair-acceptance corrections, and the long-range background subtractions are studied. The near-side jet peak widths and yields are calculated for three PV $v_\text{z}$ ranges to determine any vertex acceptance effects, with $\abs{v_\text{z}} < 3\cm$, $-15 < v_\text{z} < -3\cm$, and $3 < v_\text{z} < 15\cm$, where the nominal vertex-z range is $\abs{v_\text{z}} < 15\cm$. High-quality tracks are selected according to the criteria discussed in Section~\ref{sec:ana}. These criteria are varied to investigate their influence on the final results. The maximum allowed values for $\abs{\rd_{z}}/\sigma_{z}$ and $\abs{\rd_{xy}}/\sigma_{xy}$ are varied from 2 to 5, respectively. The maximum allowed value for $\sigma_{\pt}/\pt$ is varied from 0.05 to 0.15, and the maximum allowed normalized $\chi^{2}$ for tracks is varied from 0.15 to 0.18. The centrality calibration is varied to estimate the related systematic uncertainty in the width of the near-side peak for \PbPb collisions. The tracking efficiency is varied by its  associated uncertainty of 5\%. Pileup is negligible in \PbPb and its influence is studied in \pp collisions by determining how the number of associated tracks for reconstructed vertices varied with the distance between vertices. For the pair-acceptance correction, any residual non-uniformity observed in the $\Delta\eta$ distribution within the $\Delta\phi$ sideband region ($1.2 < \Delta\phi < \pi - 1.2$) is fitted with a smooth function, and the correlation function is flattened using this fitted shape. The effect of this correction on the near-side peak widths and yields is taken as the systematic uncertainty associated with the pair-acceptance correction. For the long-range background subtraction, two $\Delta\eta$ regions are considered to estimate the uncertainty: $2 < \abs{\Delta\eta} < 2.5$ and $\abs{\Delta\eta} > 2.5$. The near-side peak width is obtained from a double-Gaussian fit in each case, and the larger deviation between the two results is assigned as the systematic uncertainty associated with the long-range background subtraction. Individual effects for positive and negative $\Etatrig$ are considered as one of the systematic sources for longitudinal asymmetry studies. The nominal $\Delta\eta$ ranges are over 0.7 units for both $\Delta\eta > 0$ and $\Delta\eta < 0$. For the systematic variations, these ranges are changed to extend over 0.6 and 0.8 units, respectively.

Systematic uncertainties from different sources are determined after accounting for statistical fluctuations in systematic error calculations. The maximum systematic uncertainty on $\sigma_{\Delta\eta}$, $\sigma_{\Delta\phi}$ and yield ratio is within 2\%.
Table~\ref{tab:sys_label_pbpb} for \PbPb and Table~\ref{tab:sys_label_pp} for \pp summarizes the systematic uncertainties from different sources, with the total uncertainty for each observable calculated as the quadratic sum of the respective individual uncertainties.

\begin{table}[htbp]
    \topcaption{Summary of the absolute systematic uncertainties for $\sigma_{\Delta\eta}$, $\sigma_{\Delta\phi}$, and the associated yield ratio. The leftmost column lists all systematic sources, while the other columns show the range of systematic uncertainties across all \pt, $\eta$ and centrality ranges in minimum-bias \PbPb collisions. All values are scaled by a factor of $10^{3}$.}
    \vspace{\cmsTabSkip}
    \label{tab:sys_label_pbpb}
    \centering
    \renewcommand*{\arraystretch}{1.2}
    \cmsTable{
    \begin{tabular}{l c c c }
        \hline
        Systematic sources & $\sigma_{\Delta\eta}$ & $\sigma_{\Delta\phi}$ & $Y_{\Delta\eta>0}/Y_{\Delta\eta<0}$ \\
        \hline
        Vertex z & 0.016 -- 2.5 & 0.027 -- 1.3 & 4 -- 5 \\
        Track requirements & 0.12 -- 4.9 & 0.15 -- 4.9 & 4 -- 5 \\
        Centrality & 0.03 -- 1.4 & 0.02 -- 0.49 & 4 -- 5 \\
        Tracking efficiency & 0 -- 0.02 & 0 -- 0.01 & 0.33 -- 3.5 \\
        Pair acceptance correction & 0.08 -- 2.4 & 0 -- 0.85 & \NA \\
        Background subtraction & 0.53 -- 5.3 & 0.49 -- 3 & \NA \\ 
        Yield $\Delta\eta$ range & \NA & \NA & 0.98 -- 2.7 \\
        Total uncertainty & 0.85 -- 7.9 & 0.7 -- 3.1 & 8.6 -- 9.3 \\
        \hline
    \end{tabular}
    }
\end{table}

\begin{table}[htbp]
    \topcaption{Summary of the absolute systematic uncertainties for $\sigma_{\Delta\eta}$, $\sigma_{\Delta\phi}$, and the associated yield ratio. The leftmost column lists all systematic sources, while the other columns show the range of systematic uncertainties across all \pt and $\eta$ ranges in minimum-bias \pp collisions. All values are scaled by a factor of $10^{3}$.}
    \vspace{\cmsTabSkip}
    \label{tab:sys_label_pp}
    \centering
    \renewcommand*{\arraystretch}{1.2}
    \cmsTable{
    \begin{tabular}{l c c c }
        \hline
        Systematic sources & $\sigma_{\Delta\eta}$ & $\sigma_{\Delta\phi}$ & $Y_{\Delta\eta>0}/Y_{\Delta\eta<0}$ \\
        \hline
        Vertex z              & 0.036 -- 2.4         & 0 -- 1.2         & 4 -- 5 \\
        Track requirements    & 0.87 -- 2.2          & 0.35 -- 3.5         & 4 -- 5 \\
        Pileup                & 0.04 -- 0.25         & 0 -- 0.21         & \NA \\
        Tracking efficiency   & \NA          & \NA         & 0.14 -- 3.5 \\
        Pair acceptance correction     & 0.2 -- 1           &  0.014 -- 0.71                 & \NA \\
        Background subtraction & 0.31 -- 1.3 & 0.17 -- 1.5 & \NA \\
        Yield $\Delta\eta$ range & \NA             & \NA                  & 1.7 -- 2.6 \\
        Total uncertainty     & 1.6 -- 2.4           & 1 -- 3.8                  & 7 -- 7.9 \\        
        \hline
    \end{tabular}
    }
\end{table}

\section{Results}
\label{sec:res}

\begin{figure*} [htb!]
    \centering
    \includegraphics[width=0.8\textwidth]{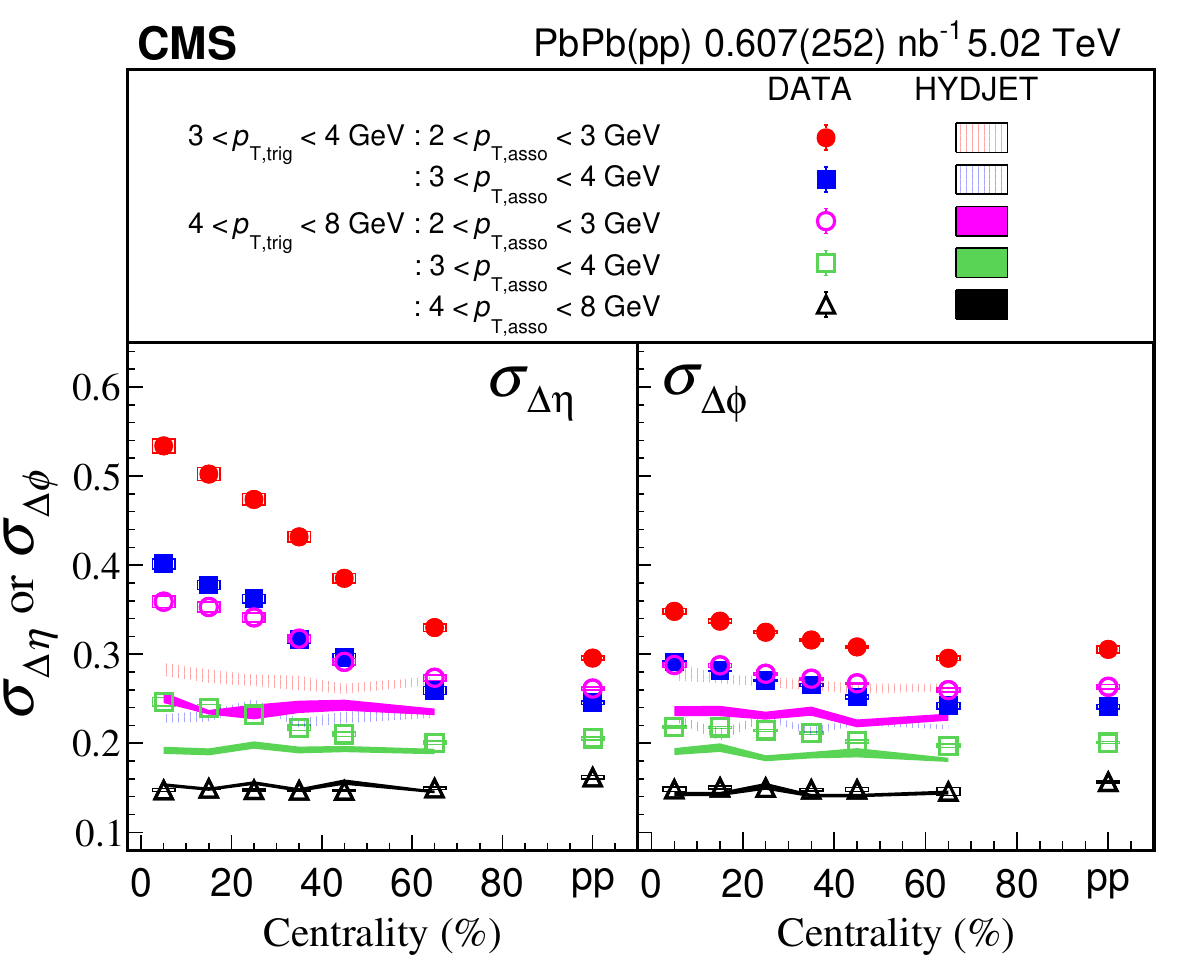}
    \caption{Centrality dependence of the longitudinal $\sigma_{\Delta\eta}$ (left panel) and transverse $\sigma_{\Delta\phi}$ (right panel) values are presented in different \pt ranges for \PbPb collisions and \pp collisions (rightmost points in each panel). The same symbols are used for the same $p_{\text{T,asso}}$ selections, with open markers representing high $p_{\text{T,trig}}$ and closed markers for low $p_{\text{T,trig}}$. The statistical uncertainties of the data points are smaller than the marker size, and rectangular boxes indicate the systematic uncertainties. Expected values from \HYDJET 1.9 are represented by bands: hatched for low $p_{\text{T,trig}}$ and solid-filled for high $p_{\text{T,trig}}$.}
    \label{fig:2D_sigma}
\end{figure*}

Figure~\ref{fig:2D_sigma} shows the centrality dependence of the longitudinal $\sigma_{\Delta\eta}$ and transverse $\sigma_{\Delta\phi}$ values in different \pt ranges of the triggered and the associated particles. It is observed that the near-side jet peak shape has similar widths in $\Delta\eta$ and $\Delta\phi$  for \pp collisions and peripheral \PbPb collisions with 50--80\% centrality. 
Both $\sigma_{\Delta\eta}$ and $\sigma_{\Delta\phi}$ increase toward central \PbPb collisions, with the longitudinal broadening in $\Delta\eta$ being significantly larger than that in $\Delta\phi$. This longitudinal broadening is most pronounced for the low-\pt trigger and associated particle ranges, while the centrality dependence of both widths becomes weaker with increasing \pt.

A possible explanation for these results is that longitudinal collective flow deforms the jet shape as the soft fragments of a jet (low-momentum hadrons that carry only a small fraction of the jet energy) interacts with the longitudinal flow of the QGP medium~\cite{Theory_flow_Jet_PRL}. Alternatively, the observed behavior could result from  energy loss of the progenitor parton of the trigger hadron in passing through the QGP medium~\cite{Theory_jet_broadening_PRL}. In contrast, \HYDJET 1.9 does not predict the centrality-dependent broadening of the longitudinal jet peak width, reflecting limitations in its modeling assumptions.

\begin{figure*} [htb!]
    \centering
    \includegraphics[width=0.8\textwidth]{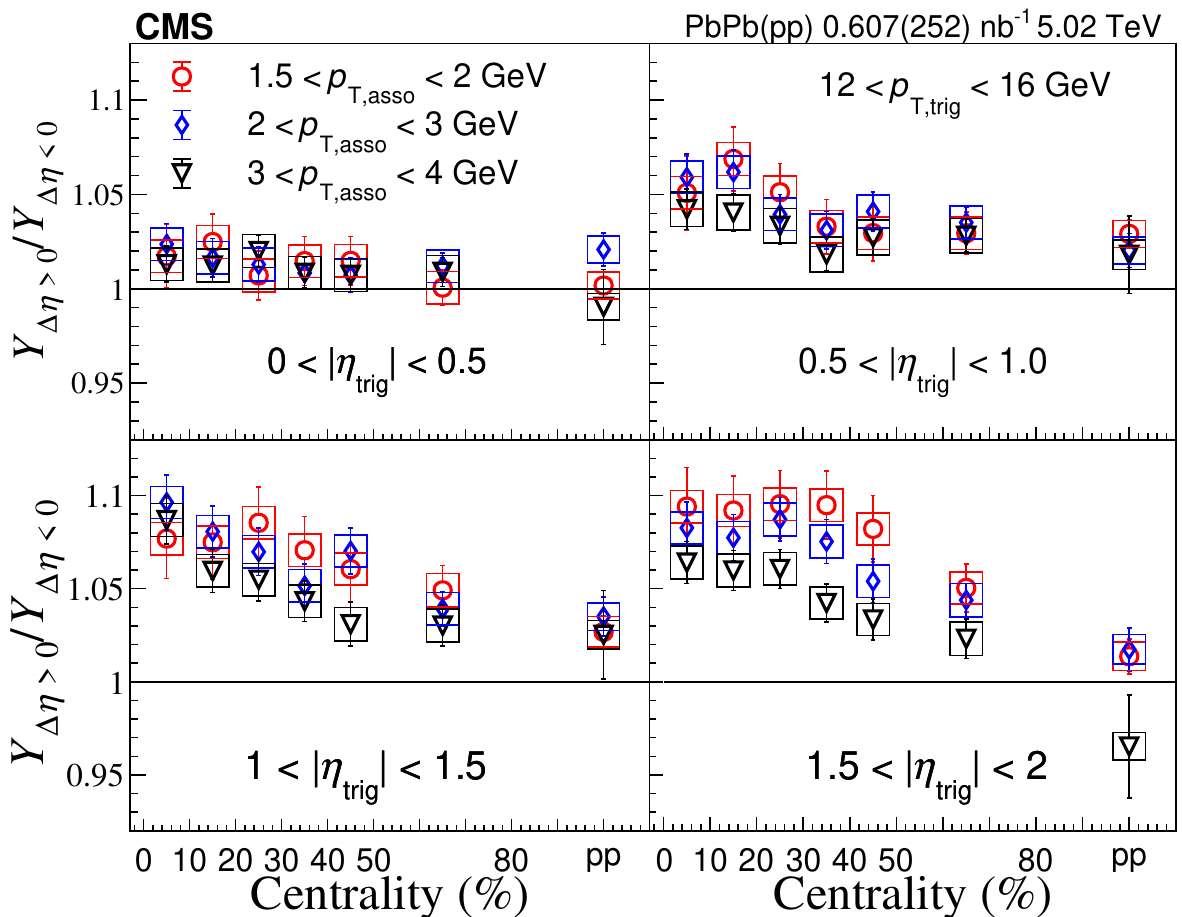}
    \caption{Associated yield ratios between $\Delta\eta > 0$ and $\Delta\eta < 0$ within the range of $0 < \abs{\Delta\eta} < 0.7$ as a function of centrality in different $\pTasso$ and $\Etatrig$ ranges for \PbPb and \pp collisions (rightmost bin) are presented in the above panel. The above panel consists of four plots, each corresponding to a different value of $\Etatrig$. Each plot consists of three colored markers representing different $\pTasso$ values ($1.5 < \pTasso < 2\GeV$, $2 < \pTasso < 3\GeV$, $3 < \pTasso < 4\GeV$), while $\pTtrig$ is fixed ($12 < \pTtrig < 16\GeV$). The vertical bars indicate statistical uncertainties, while the rectangular open boxes represent systematic uncertainties.}
    \label{fig:yield_ratio}
\end{figure*}

Figure~\ref{fig:yield_ratio} shows the associated yield ratio $Y_{\Delta\eta>0}/Y_{\Delta\eta<0}$ as a function of centrality for different $p_\text{T}$ ranges of associated particles. The measurement is also presented in different $\eta$ ranges of the triggered particles, arranged from top left to bottom right. The \pp results are shown in the last bin of each plot. 

A significant increase in the associated yield ratio is observed as the selection moves toward forward pseudorapidity (high $\Etatrig$). At midrapidity (low $\Etatrig$), the associated yield ratio remains close to unity and shows minimal dependence on $\pTasso$, indicating a nearly symmetric near-side peak structure. However, at larger $\Etatrig$, the ratio exhibits a significant increase, particularly for central \PbPb collisions, along with a pronounced dependence on $\pTasso$. 

This asymmetry trend suggests a systematic shift in the jet-like correlation structure toward positive $\Delta\eta$ as the $\eta$ of the trigger particle increases. One possible interpretation of this $\eta$ dependence of the near-side jet peak asymmetry, manifested as a higher associated yield for $\Delta\eta > 0$ (where $\Etatrig$ is always positive by construction), is that the soft fragments of a high-\pt jet experience a forward-directed push resulting from interactions with the longitudinally expanding medium~\cite{Theory_flow_Jet_PRL}. The increasing asymmetry with collision centrality further supports the idea that medium effects play a significant role. This asymmetric behavior suggests a potential breaking of the longitudinal invariance of the near-side peak across the $\Delta\eta$ range.

\section{Summary}
\label{sec:sum}

The centrality and pseudorapidity ($\eta$) dependencies of jet peak shapes have been explored using two-particle correlations. Lead-lead (\PbPb) collision data at a center-of-mass energy per nucleon pair of 5.02\TeV were recorded using the CMS detector at the CERN LHC. These minimum-bias data correspond to an integrated luminosity of 0.607\nbinv. Proton-proton (\pp) collision data at the same nucleon-nucleon collision energy and with an integrated luminosity of 252\nbinv are also analyzed to provide a reference. The separations in pseudorapidity ($\Delta\eta$) and azimuth ($\Delta\phi$) of all particles within specific transverse momentum (\pt) ranges are used to form two-dimensional correlation distributions that are averaged over all events. Jets result in a peak shaped structure corresponding to particle pairs with small angular separation. Jet peak shapes, characterized by associated widths in $\Delta\eta$ and $\Delta\phi$, are presented as functions of centrality. The jet peak asymmetry as a function of $\eta$ is also studied. 

For \PbPb collisions, both the longitudinal ($\sigma_{\Delta\eta}$) and transverse ($\sigma_{\Delta\phi}$) widths of the near-side jet peak exhibit a centrality-dependent broadening compared to the \pp reference. The effect is more pronounced in $\Delta\eta$, indicating a stronger modification of the correlation structure along the longitudinal direction. At higher particle transverse momenta, the centrality dependence of both widths becomes weaker, approaching the \pp baseline within uncertainties. Overall, the stronger longitudinal than transverse broadening may reflect the influence of longitudinal expansion or partonic energy loss mechanisms in the medium.
 
The asymmetry of the jet peak distribution in $\Delta\eta$ is studied through the ratio of yields on the positive and negative sides of the jet peak. This ratio is evaluated in distinct $\abs{\Etatrig}$ ranges and for various energy thresholds of the higher-\pt particle in each pair. The ratio is found to increase with the average $\abs{\Etatrig}$ value. For a given $\abs{\Etatrig}$ range, the ratio measured in peripheral events is found similar to that in \pp collisions, but increases significantly towards more central collisions. These results suggest that the jet properties are modified through in medium interactions in a longitudinally expanding QGP medium.

\begin{acknowledgments}
We congratulate our colleagues in the CERN accelerator departments for the excellent performance of the LHC and thank the technical and administrative staffs at CERN and at other CMS institutes for their contributions to the success of the CMS effort. In addition, we gratefully acknowledge the computing centers and personnel of the Worldwide LHC Computing Grid and other centers for delivering so effectively the computing infrastructure essential to our analyses. Finally, we acknowledge the enduring support for the construction and operation of the LHC, the CMS detector, and the supporting computing infrastructure provided by the following funding agencies: SC (Armenia), BMBWF and FWF (Austria); FNRS and FWO (Belgium); CNPq, CAPES, FAPERJ, FAPERGS, and FAPESP (Brazil); MES and BNSF (Bulgaria); CERN; CAS, MoST, and NSFC (China); MINCIENCIAS (Colombia); MSES and CSF (Croatia); RIF (Cyprus); SENESCYT (Ecuador); ERC PRG and PSG, TARISTU24-TK10 and MoER TK202 (Estonia); Academy of Finland, MEC, and HIP (Finland); CEA and CNRS/IN2P3 (France); SRNSF (Georgia); BMFTR, DFG, and HGF (Germany); GSRI (Greece); NKFIH (Hungary); DAE and DST (India); IPM (Iran); SFI (Ireland); INFN (Italy); MSIT and NRF (Republic of Korea); MES (Latvia); LMTLT (Lithuania); MOE and UM (Malaysia); BUAP, CINVESTAV, CONACYT, LNS, SEP, and UASLP-FAI (Mexico); MOS (Montenegro); MBIE (New Zealand); PAEC (Pakistan); MES, NSC, and NAWA (Poland); FCT (Portugal); MESTD (Serbia); MICIU/AEI and PCTI (Spain); MOSTR (Sri Lanka); Swiss Funding Agencies (Switzerland); MST (Taipei); MHESI (Thailand); TUBITAK and TENMAK (T\"{u}rkiye); NASU (Ukraine); STFC (United Kingdom); DOE and NSF (USA).

\hyphenation{Rachada-pisek} Individuals have received support from the Marie-Curie program and the European Research Council and Horizon 2020 Grant, contract Nos.\ 675440, 724704, 752730, 758316, 765710, 824093, 101115353, 101002207, 101001205, and COST Action CA16108 (European Union); the Leventis Foundation; the Alfred P.\ Sloan Foundation; the Alexander von Humboldt Foundation; the Science Committee, project no. 22rl-037 (Armenia); the Fonds pour la Formation \`a la Recherche dans l'Industrie et dans l'Agriculture (FRIA) and Fonds voor Wetenschappelijk Onderzoek contract No. 1228724N (Belgium); the Beijing Municipal Science \& Technology Commission, No. Z191100007219010, the Fundamental Research Funds for the Central Universities, the Ministry of Science and Technology of China under Grant No. 2023YFA1605804, the Natural Science Foundation of China under Grant No. 12535004, and USTC Research Funds of the Double First-Class Initiative No.\ YD2030002017 (China); the Ministry of Education, Youth and Sports (MEYS) of the Czech Republic; the Shota Rustaveli National Science Foundation, grant FR-22-985 (Georgia); the Deutsche Forschungsgemeinschaft (DFG), among others, under Germany's Excellence Strategy -- EXC 2121 ``Quantum Universe" -- 390833306, and under project number 400140256 - GRK2497; the Hellenic Foundation for Research and Innovation (HFRI), Project Number 2288 (Greece); the Hungarian Academy of Sciences, the New National Excellence Program - \'UNKP, the NKFIH research grants K 131991, K 133046, K 138136, K 143460, K 143477, K 146913, K 146914, K 147048, 2020-2.2.1-ED-2021-00181, TKP2021-NKTA-64, and 2025-1.1.5-NEMZ\_KI-2025-00004 (Hungary); the Council of Science and Industrial Research, India; ICSC -- National Research Center for High Performance Computing, Big Data and Quantum Computing, FAIR -- Future Artificial Intelligence Research, and CUP I53D23001070006 (Mission 4 Component 1), funded by the NextGenerationEU program, the Italian Ministry of University and Research (MUR) under Bando PRIN 2022 -- CUP I53C24002390006, PRIN PRIMULA 2022RBYK7T (Italy); the Latvian Council of Science; the Ministry of Education and Science, project no. 2022/WK/14, and the National Science Center, contracts Opus 2021/41/B/ST2/01369, 2021/43/B/ST2/01552, 2023/49/B/ST2/03273, and the NAWA contract BPN/PPO/2021/1/00011 (Poland); the Funda\c{c}\~ao para a Ci\^encia e a Tecnologia (Portugal); the National Priorities Research Program by Qatar National Research Fund; MICIU/AEI/10.13039/501100011033, ERDF/EU, ``European Union NextGenerationEU/PRTR", projects PID2022-142604OB-C21, PID2022-139519OB-C21, PID2023-147706NB-I00, PID2023-148896NB-I00, PID2023-146983NB-I00, PID2023-147115NB-I00, PID2023-148418NB-C41, PID2023-148418NB-C42, PID2023-148418NB-C43, PID2023-148418NB-C44, PID2024-158190NB-C22, RYC2021-033305-I, RYC2024-048719-I, CNS2023-144781, CNS2024-154769 and Plan de Ciencia, Tecnolog{\'i}a e Innovaci{\'o}n de Asturias, Spain; the Chulalongkorn Academic into Its 2nd Century Project Advancement Project, the National Science, Research and Innovation Fund program IND\_FF\_68\_369\_2300\_097, and the Program Management Unit for Human Resources \& Institutional Development, Research and Innovation, grant B39G680009 (Thailand); the Eric \& Wendy Schmidt Fund for Strategic Innovation through the CERN Next Generation Triggers project under grant agreement number SIF-2023-004; the Kavli Foundation; the Nvidia Corporation; the SuperMicro Corporation; the Welch Foundation, contract C-1845; and the Weston Havens Foundation (USA).
\end{acknowledgments}\section*{Data availability} Release and preservation of data used by the CMS Collaboration as the basis for publications is guided by the  \href{https://doi.org/10.7483/OPENDATA.CMS.1BNU.8V1W}{CMS data preservation, re-use and open access policy}.

\bibliography{auto_generated} 
\clearpage

\cleardoublepage \appendix\section{The CMS Collaboration \label{app:collab}}\begin{sloppypar}\hyphenpenalty=5000\widowpenalty=500\clubpenalty=5000\cmsinstitute{Yerevan Physics Institute, Yerevan, Armenia}
{\tolerance=6000
A.~Belyaev\cmsorcid{0000-0003-1692-1173}, A.~Hayrapetyan, V.~Makarenko\cmsorcid{0000-0002-8406-8605}, A.~Tumasyan\cmsAuthorMark{1}\cmsorcid{0009-0000-0684-6742}
\par}
\cmsinstitute{Institut f\"{u}r Hochenergiephysik, Vienna, Austria}
{\tolerance=6000
W.~Adam\cmsorcid{0000-0001-9099-4341}, J.W.~Andrejkovic, L.~Benato\cmsorcid{0000-0001-5135-7489}, T.~Bergauer\cmsorcid{0000-0002-5786-0293}, K.~Damanakis\cmsorcid{0000-0001-5389-2872}, M.~Dragicevic\cmsorcid{0000-0003-1967-6783}, C.~Giordano\cmsorcid{0000-0001-6317-2481}, P.S.~Hussain\cmsorcid{0000-0002-4825-5278}, M.~Jeitler\cmsAuthorMark{2}\cmsorcid{0000-0002-5141-9560}, N.~Krammer\cmsorcid{0000-0002-0548-0985}, A.~Li\cmsorcid{0000-0002-4547-116X}, D.~Liko\cmsorcid{0000-0002-3380-473X}, I.~Mikulec\cmsorcid{0000-0003-0385-2746}, J.~Schieck\cmsAuthorMark{2}\cmsorcid{0000-0002-1058-8093}, R.~Sch\"{o}fbeck\cmsAuthorMark{2}\cmsorcid{0000-0002-2332-8784}, D.~Schwarz\cmsorcid{0000-0002-3821-7331}, M.~Shooshtari\cmsorcid{0009-0004-8882-4887}, M.~Sonawane\cmsorcid{0000-0003-0510-7010}, W.~Waltenberger\cmsorcid{0000-0002-6215-7228}, C.-E.~Wulz\cmsAuthorMark{2}\cmsorcid{0000-0001-9226-5812}
% ...
\par}
\vskip\cmsinstskip
\dag:~Deceased\\
$^{1}$Also at Yerevan State University, Yerevan, Armenia\\
$^{2}$Also at TU Wien, Vienna, Austria\\
% ...
\end{sloppypar}
\end{document}